\documentclass[5p, sort&compress]{elsarticle} 
\usepackage{graphicx}
\usepackage{amssymb}
\usepackage{epstopdf}
\usepackage[amssymb,thinqspace,thinspace]{SIunits}
\usepackage{amsmath}
\usepackage{booktabs}
\usepackage{rotating}
\usepackage{multirow}
\usepackage{psfrag}
\usepackage{adjustbox}
\usepackage{csquotes}
\usepackage{arydshln}
\usepackage{textgreek}
\usepackage{color}
\usepackage{ulem}
\usepackage[super]{nth}
\usepackage{comment}
\usepackage{dblfloatfix}
\usepackage{url} 

\journal{arXiv}

\begin{document}
\begin{frontmatter}
\title{The mechanism of twin thickening and the elastic strain state of TWIP steel nanotwins}

\author[IC]{TWJ Kwok}
\author[IC]{TP McAuliffe}
\author[IC]{AK Ackerman}
\author[NCEM]{BH Savitzky}
\author[EPSIC]{M Danaie}
\author[NCEM]{C Ophus}
\author[IC]{D Dye\corref{cor1}}\ead{david.dye@imperial.ac.uk}
\cortext[cor1]{Corresponding author}
\address[IC]{Department of Materials, Royal School of Mines, Imperial College London, Prince Consort Road, London, UK}
\address[NCEM]{National Center for Electron Microscopy (NCEM), Molecular Foundry, Lawrence Berkeley Lab, USA}
\address[EPSIC]{Electron Physical Science Imaging Centre (ePSIC), Diamond Light Source, Oxford, UK}

\begin{abstract}

A Twinning Induced Plasticity (TWIP) steel with a nominal composition of Fe-16.4Mn-0.9C-0.5Si-0.05Nb-0.05V was deformed to an engineering strain of 6\%. The strain around the deformation twins were mapped using the 4D-STEM technique. Strain mapping showed a large average elastic strain of approximately 6\% in the directions parallel and perpendicular to the twinning direction. However, the large average strain comprised of several hot spots of even larger strains of up to 12\%. These hot spots could be attributed to a high density of sessile Frank dislocations on the twin boundary and correspond to shear stresses of 1--1.5 GPa. The strain and therefore stress fields are significantly larger than other materials known to twin and are speculated to be responsible for the early thickness saturation of TWIP steel nanotwins. The ability to keep twins extremely thin helps improve grain fragmentation, \textit{i.e.} the dynamic Hall-Petch effect, and underpins the large elongations and strain hardening rates in TWIP steels.




\end{abstract}

\end{frontmatter}



Twinning has become increasingly significant in modern metallurgy \cite{Lu2009}. Deformation or mechanical twinning occurs when a region of the crystal is sheared, with each atom moving less than one lattice spacing, such that the twinned region forms a mirror image of the parent crystal about the twinning or habit plane \cite{Cottrell1995}. Many modern materials such as TWIP steels \cite{Rahman2014,Rahman2015a}, medium Mn steels \cite{Kwok2022e}, TWIP Ti \cite{Zhao2020a,Gao2018}, Mg alloys and Zr alloys \cite{Abdolvand2016} all capitalise on twinning to enable high strain hardening rates and improve elongation. However, as Qin and Bhadeshia \cite{Qin2008} point out, the maximum contribution of twinning to total elongation (true strain) in a random polycrystal is $<0.3$. This is compared to total elongation of 0.5--0.6 true strain in TWIP steels \cite{DeCooman2018,DeCooman2011}. It is therefore the interaction between twins and dislocations which are of importance. In TWIP steels, long and thin twins are formed during deformation, acting as barriers to dislocation motion. The volume fraction of twins increase with increasing strain and continuously subdivides the grain, resulting in a high strain hardening rate and also elongation. This mechanism has been termed the dynamic Hall-Petch effect \cite{Allain2004a,Bouaziz2001}.

\begin{figure*}[t]
	\centering
	\includegraphics[width=\linewidth]{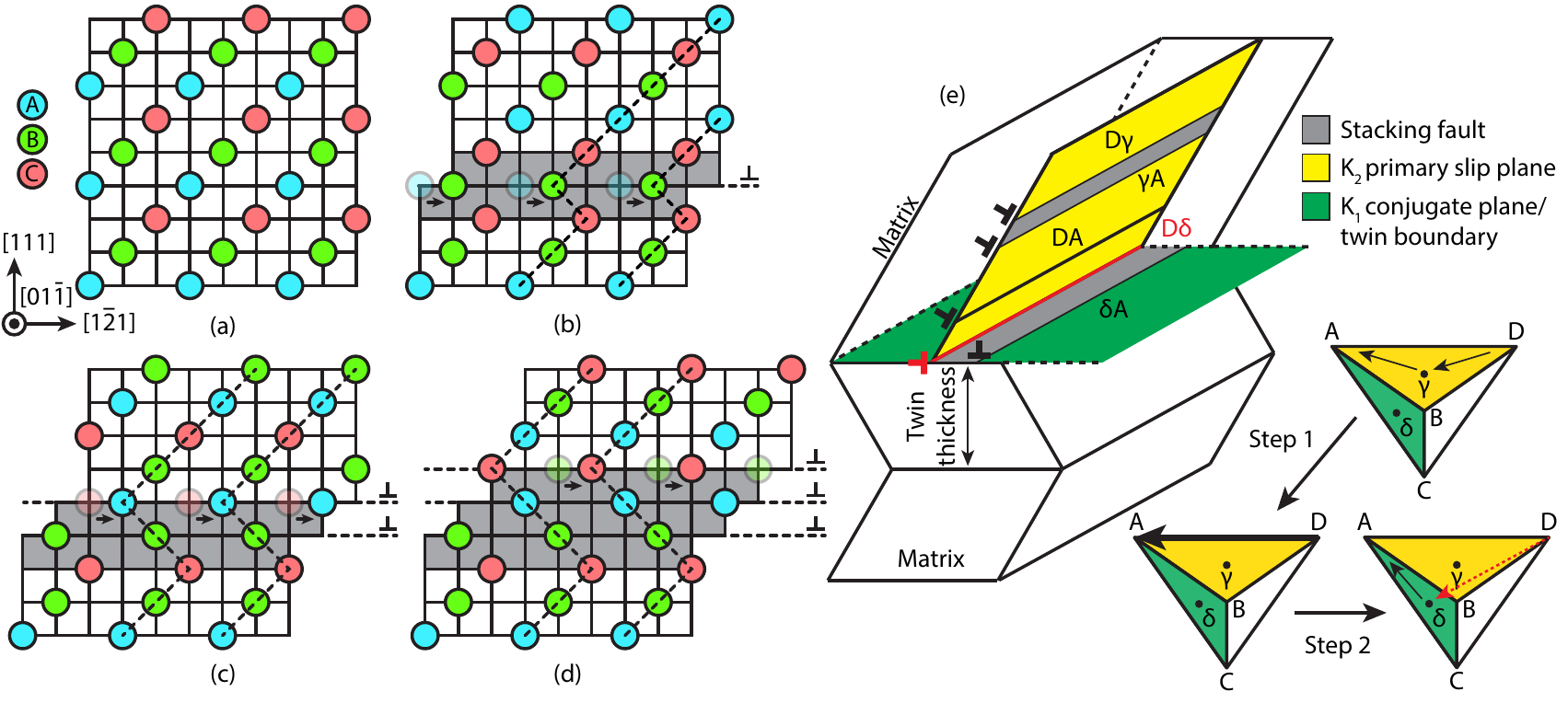}
	\caption{Schematic ball and stick model of twinning in FCC materials. (a) Perfect FCC lattice with ABCABC stacking. (b) Passage of one, (c) two and (d) three Shockley partial dislocations on successive (111) planes building up the twin thickness. Grey regions indicate the growing twin. (e) Schematic of the interaction between dislocation and twin boundaries described by Idrissi and Schryvers \cite{Idrissi2012}, drawn after De Cooman \textit{et al.} \cite{DeCooman2018}.}
	\label{fig:atom-twins}
\end{figure*}

In the case of Face Centred Cubic (FCC) crystals, the origin of twinning is still a matter of academic debate with many different mechanisms being proposed although the three layer twin nucleus model by Mahajan and Chin \cite{Mahajan1973} is increasingly being used in computational models \cite{Steinmetz2013,DeCooman2018}. Nevertheless, the fundamentals of FCC twinning are well understood and first involves the dissociation of a full $\frac{a}{2} \{111\} \langle 110\rangle$ into two $\frac{a}{6} \{111\} \langle112\rangle$ Shockley partial dislocations, where $a$ is the FCC lattice spacing. In TWIP steels, the formation of partial dislocations is favoured due to the low Stacking Fault Energy (SFE). An FCC twin is then created, as shown in Figure \ref{fig:atom-twins}, when partial dislocations propagate on successive $\{111\}$ planes and increasing the twin thickness. 

In TWIP steels, the deformation twins are often between 20--30 nm in thickness, corresponding to the glide of approximately 100 partial dislocations \cite{DeCooman2018}. Interestingly, these twins generally do not continue to thicken with strain once this thickness is reached, rather choosing to form new twins instead. This is compared to a larger twin thickness of $\leq100$ nm in Cu-Al alloys \cite{Zhang2009} and 1--2 \textmu m in $\alpha$-brass \cite{Roy2014}. Therefore, while there is a general mechanism for twin thickening, there also appears to be mechanism for twin thickness saturation based on some material property such as SFE \cite{Zhang2009} and grain size \cite{Rahman2015,Ghaderi2011}. Unfortunately, the reasons for twin thickness saturation are not very well studied, even though the ability of TWIP steels to form a high density of very fine twins is paramount to a high degree of grain fragmentation, underpinning the dynamic Hall-Petch effect and overall tensile properties.

Nevertheless, Idrissi and Schryvers \cite{Idrissi2012} studied the interaction between coherent twin boundaries and dislocations and proposed a \enquote{pole $+$ deviation} mechanism that might help to control twin thickness. Illustrated in Figure \ref{fig:atom-twins}e, the first step involves two Shockley partials (D$\gamma$ and $\gamma$A) gliding on the K\textsubscript{1} primary slip plane which constrict and recombine under applied stress to form a full dislocation (DA). In the second step, the full dislocation meets the coherent twin boundary and dissociates again under stress into a sessile Frank partial dislocation (D$\delta$) and a Shockley partial dislocation ($\delta$A) according to the reaction:

\begin{equation}
	\frac{a}{2}\langle110\rangle \rightarrow \frac{a}{3}\langle111\rangle_{sessile} + \frac{a}{6}\langle11\bar{2}\rangle
\end{equation}

\noindent
where the sessile Frank partial dislocation (D$\delta$) remains along the twin boundary and the Shockley partial dislocation ($\delta$A) glides off on the K\textsubscript{2} conjugate plane, increasing the twin thickness by one layer. Idrissi and Schryvers \cite{Idrissi2012} then propose that the high density of sessile Frank partials on the twin boundary will affect the mobility of Shockley partials on the twinning plane (K\textsubscript{1}), limiting the thickening of the twin. Recent computational work by Grilli \textit{et al.} \cite{Grilli2020} confirmed the observations by Idrissi and Schryvers \cite{Idrissi2012} that residual dislocations on the twin boundary can completely stop twin thickening and add that materials with a larger SFE produces larger twins.

To further investigate the reasons behind the early twin thickness saturation in TWIP steels, Four Dimensional Scanning Transmission Electron Microscopy (4D-STEM) can used to map the strain state in and around a mechanical twin. 4D-STEM is a relatively new technique where an electron diffraction pattern is acquired at every point in a scan grid where sub-nm spatial resolution can be achieved \cite{Ophus2019}. The term \enquote{4D} refers to the collection of 2D diffraction patterns over a 2D grid. A comprehensive review of 4D-STEM and its applications in strain mapping, imaging and ptychography is available in a review by Ophus \cite{Ophus2019}.



In this study, a 400 g ingot of steel with nominal composition of Fe-16.4Mn-0.9C-0.5Si-0.05Nb-0.05V (SFE $\approx$ 23 mJ m\textsuperscript{-2} \cite{Sun2018}) based on the composition by Scott \textit{et al.} \cite{Scott2011} was produced by arc melting pure elements in a vacuum arc melter. The steel was cast into a copper crucible and homogenised at 1300 \degree C for 24 h in Ar. The steel ingot was then hot rolled from $10 \rightarrow 5$ mm (50\% reduction) in 2 passes at 1000 \degree C, cold rolled from $5 \rightarrow 1.5$ mm (67\% reduction) and finally annealed at 1000 \degree C for 5 min. Sub-sized tensile samples with gauge dimensions of 19 $\times$ 1.5 $\times$ 1.5 mm were obtained from the rolled strip \textit{via} electrical discharge machining. Tensile samples were then tested to 6\% engineering strain (approximately 5\% plastic strain) and to failure, both at a nominal strain rate of $10^{-3}$ s\textsuperscript{-1}.

Electron Backscattered Diffraction (EBSD) was conducted using a Sigma 300 FEG-SEM on the as-annealed and 6\% strained conditions. A strain of 6\% was chosen to produce a sufficient twin density in order to locate the twins using EBSD while keeping background dislocation density to a minimum. In the 6\% strained sample, a grain with a $\langle111\rangle$ crystallographic direction in the vertical axis of the microscope field of view was selected and a thin foil was extracted using the Focussed Ion Beam (FIB) lift-out technique, producing a foil with its $\langle111\rangle$ direction parallel to the foil normal.


4D-STEM was performed on the foil using the probe corrected JEOL ARM200CF Transmission Electron Microscope (TEM) at ePSIC (Oxford, United Kingdom). Diffraction patterns were collected with a Merlin (MediPix) direct electron detector. An accelerating voltage of 200 kV and a camera length of 40 cm were employed with a 10 \textmu m CL1 aperture. A $68.2 \times 83.1$ nm area of interest was scanned in $188 \times 229$ real space pixels and with a $256 \times 256$ pixel diffraction pattern captured at each of these scan positions with a 1 ms dwell time per pattern.

The in-plane elastic strain was calculated from the electron diffraction pattern at each scan location. Bragg peak identification, dataset calibration including elliptical distortion correction and diffraction shift correction and elastic strain calculation were performed with the open source py4DSTEM analysis package \cite{Savitzky2019}. For locating Bragg peaks, a correlation power of 1 was used, corresponding to cross-correlation \cite{Pekin2017}, and achieve subpixel precision with local Fourier up-sampling by a factor of 16 \cite{Guizar-Sicairos2008}. For the rest of the paper, an orthogonal frame of reference with directions \textit{1, 2, 3} will be adopted, where \textit{3} refers to the $[111]$ direction perpendicular to the surface of the foil, \textit{1} as $[1\bar{2}1]$ and \textit{2} as $[10\bar1]$.

\begin{figure}[t!]
	\centering
	\includegraphics[width=\linewidth]{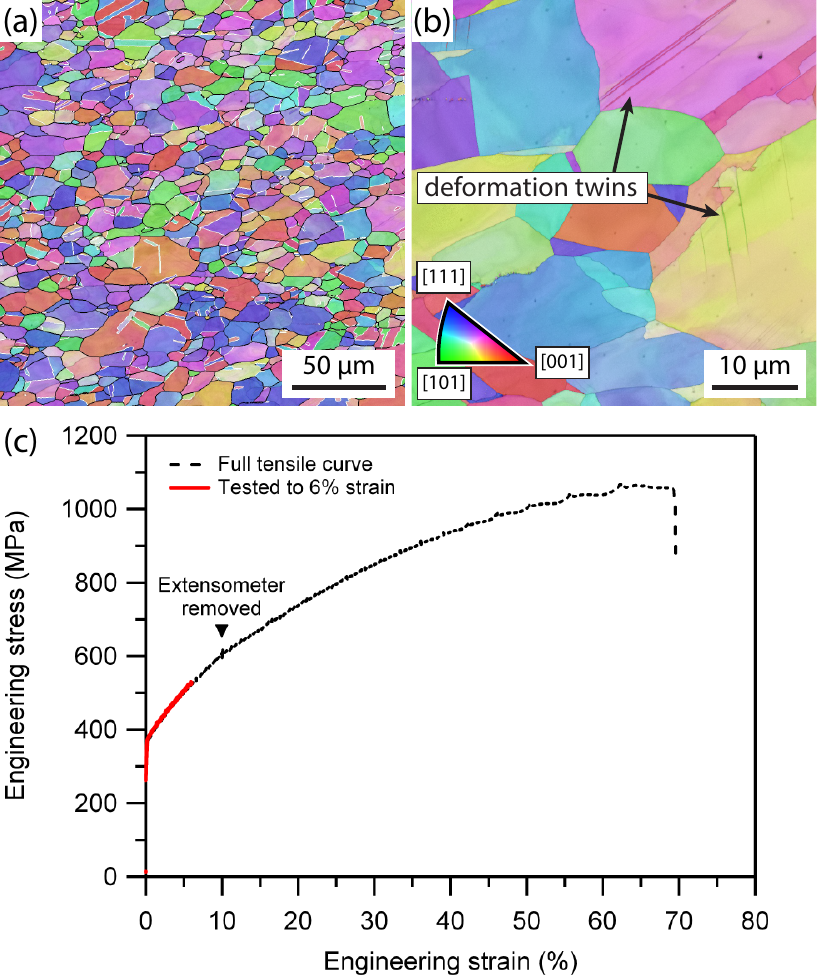}
	\caption{EBSD IPF-X (right) and image quality maps of (a) as-annealed sample where black lines indicate high angle grain boundaries and white lines indicate austenite $\Sigma3$ boundaries (rolling direction parallel to the horizontal axis and transverse direction pointing out of the page), (b) sample deformed to 6\% strain, demonstrating fine deformation twins. (c) Engineering tensile curves of the investigated steel when tested to failure and to 6\% strain. }
	\label{fig:tensile-ebsd}
\end{figure}


The microstructure of the as-annealed TWIP steel is shown in Figure \ref{fig:tensile-ebsd}a with an area weighted average grain size of 12 \textmu m. An EBSD map of the 6\% strained sample is shown in Figure \ref{fig:tensile-ebsd}b. Unfortunately, due to the fine twin thickness, it is not easy to index deformation twins using EBSD. However, they can be identified as darker lines in the Image Quality (IQ) map. The tensile behaviour of the investigated steel is shown in Figure \ref{fig:tensile-ebsd}c and has a yield strength of 360 MPa, tensile strength of 1050 MPa and total elongation of 69\%.

\begin{figure*}[t]
	\centering
	\includegraphics{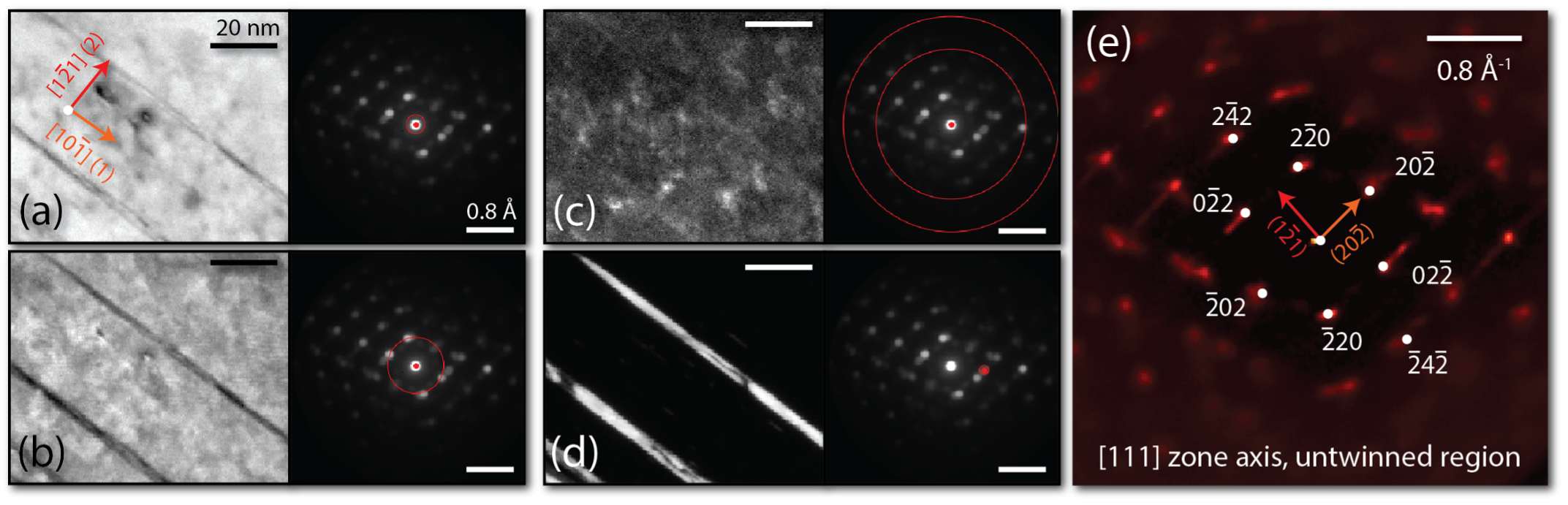}
	\caption{(a,b) Virtual bright and (c,d) dark field images of the steel nanotwin using different virtual apertures indicated by the open red circles. (e) Average deconvolution maps, sums over real space of the identified Bragg peak locations weighted by intensity.}
	\label{fig:vbf-vdf}
\end{figure*}

Figure \ref{fig:vbf-vdf}a--d shows the Virtual Bright and Dark Field (VBF and VDF respectively) images reconstructed from the intensity collected from the highlighted digital apertures. VBF images in Figure \ref{fig:vbf-vdf}a--b clearly distinguish the twin from the matrix. Given the axis system we have adopted with [111] out of the plane and the diffraction vectors indexed in Figure \ref{fig:vbf-vdf}e, it is inferred that the twins have a habit plane of $(11\bar{1})$.

In conventional TEM, the direct beam contains a large amount of structure-dependent information. Conventional bright field imaging when used with an objective aperture to isolate the direct beam, uses electron wave phase information as well as intensity to re-interfere and reconstruct an image \cite{Williams2009}. In VBF, only access to electron intensity in the diffraction plane is available. Therefore it is likely that the observed contrast between twin and matrix is derived from local strain, lattice rotation or dynamical effects which will alter the ratio of diffracted to direct intensity. The high angle VDF image in Figure \ref{fig:vbf-vdf}c is similar to convention High-Angle Annular Dark Field (HAADF) imaging which detects electrons which are scattered to high angles and also suggests that there was no detectable variation in local chemistry between the twin and the matrix. The twin is explicitly highlighted in Figure \ref{fig:vbf-vdf}d by reconstructing the spatial image from the $00\bar{2}$ reciprocal lattice point for the twinned region only, similar to a conventional TEM dark field image. The action of twinning rotates the the diffraction pattern about the direct beam, such that the twinned and untwinned $02\bar{2}$ peaks are separated. In Figure \ref{fig:vbf-vdf}e, a Bragg vector map is presented (after Savitzky \textit{et al.} \cite{Savitzky2019}), showing the distribution of all measured Bragg peak locations for the untwinned region. It is acknowledged that a small amount of intensity is observed at the twin reciprocal lattice points even for the untwinned material. This is possibly due to the geometry of the sample, \textit{i.e.} twinning habit plane not perfectly parallel to beam direction, and through thickness sampling of both untwinned and twinned material near the interface. 

\begin{figure*}[t!]
	\centering
	\includegraphics[width=\linewidth]{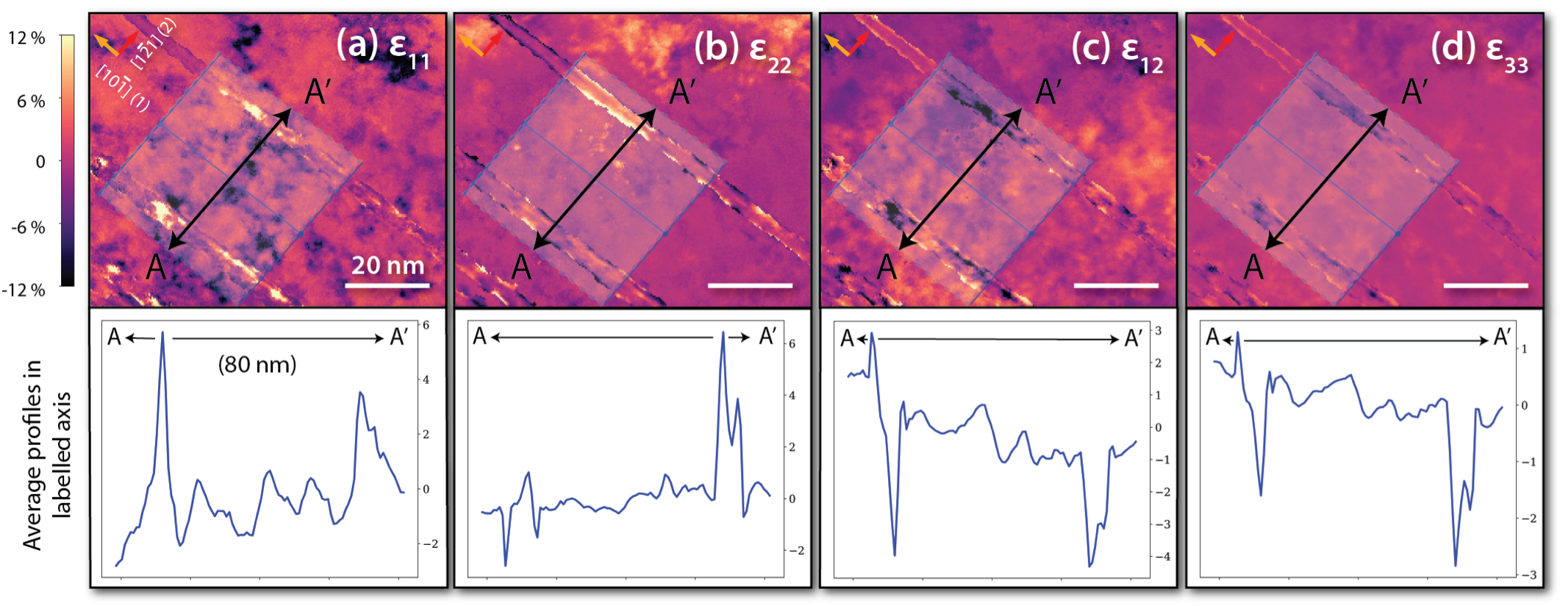}
	\caption{Elastic strain maps resolved in the (a) 11, (b) 22, (c) 12 and (d) 33 directions. The average 1-direction profile, perpendicular to the twin’s length, was calculated by integrating all points along the \textit{2}-direction within the highlighted region.}
	\label{fig:twinstrain}
\end{figure*}

The strain components $\varepsilon_{11}$, $\varepsilon_{22}$, $\varepsilon_{12}$ were calculated from the relative movements of the diffraction spots. This operation was performed independently for twinned and untwinned regions. A set of reference reciprocal lattice vectors were obtained by averaging the untwinned regions's reciprocal lattice basis vectors. The magnitude of the twin basis vectors were normalised to this unstrained length. As such, elastic strains are given with reference to this \enquote{unstrained} state. The measurement could alternatively be considered as the elastic strain variation across the area of interest. 


Maps of the measured (11, 22, 12) and inferred strain components across the area of interest are shown in Figure \ref{fig:twinstrain}. An average line profile in the 1-direction was obtained by integrating points along the 2-direction within the highlighted area. From the average profiles in Figure \ref{fig:twinstrain}, it is striking that the average elastic strains in the 11 and 22 directions reach values as high as 6\% at locations along the twin boundaries. Hot spots of up to 12\% were also observed at certain points along the twin boundaries. These values far exceed other elastic strain measurements observed in the literature (Table \ref{tab:straincompare}), even between other 4D-STEM experiments with the same methodology, \textit{e.g.} $\sim4\%$ observed by Pekin \textit{et al.} along twin boundaries in stainless steel 304 \cite{Pekin2017}. However, it should be noted that the averaged strains have been disproportionately raised above the baseline of approximately 4\% by the presence of hot spots. 

\begin{table}[t]
	\centering
	\caption{Comparison of strain, $\varepsilon$, in \% obtained with different strain mapping methods on various materials. *SS--Stainless Steel, CP--Commercially Pure, N-PED--Nano-Precession Electron Diffraction.}
	\begin{adjustbox}{width=\columnwidth,center}
		\begin{tabular}{lcccc}
			\toprule
			Material & Method & Feature & $\varepsilon$ & Reference \\
			\midrule
			
			AISI 321 (SS)	&	4D-STEM	&	Planar dislocations	&	0.5	& \cite{Pekin2018} \\
			Boron Carbide & N-PED & Grain boundary &  0.75     & \cite{Rottmann2018} \\
			Pd    & HR-TEM & Twins & 2.5   & \cite{Rosner2010} \\
			SS 304 & 4D-STEM  & Twins & 4     & \cite{Pekin2017} \\
			CP Mg & 4D-STEM  & Twins & 4     & \cite{Chen2020b} \\
			TWIP steel & 4D-STEM & Twins & 6     & Current \\

			\bottomrule
		\end{tabular}%
	\end{adjustbox}
	\label{tab:straincompare}%
\end{table}%

The presence of hot spots was also observed by R\"{o}sner \textit{et al.} \cite{Rosner2010} in a cold rolled Pd foil. These hot spots of approximately $\leq25\%$ strain were attributed to dislocation debris when mapping the strain of dislocation cores of stacking faults or partial dislocations in the foil. It should therefore also follow, that the hot spots observed at the twin boundaries in Figure \ref{fig:twinstrain} were the result of a highly dislocated structure. Following the \enquote{pole $+$ deviation} mechanism proposed by Idrissi and Schryvers \cite{Idrissi2012}, the hot spots are likely to be the locations where one slip system impinges onto the twin boundary, generating a high density of sessile Frank partial dislocations at the twin boundary and also generating a very large stress field. Faulted twin boundaries as a result of impingement from another slip system were also shown by Dao \textit{et al.} \cite{Dao2006} in nanotwinned Cu using HR-TEM. Furthermore, since the sample was only deformed to 5\% strain, it is likely that only a few slip systems would have been activated, hence the localisation of strain only at certain points along the entire twin boundary. 

In the study by Grilli \textit{et al.} \cite{Grilli2020}, it was found that an additional shear stress of 400 MPa to the critical resolved shear stress for twinning in $\alpha$-uranium (orthorhombic) was sufficient to completely retard twinning and favour slip instead. Using the analytical model from M\"{u}llner \textit{et al.} \cite{Mullner1994,Mullner1997,Mullner1994a}, it was estimated that the strain hot spots at the twin boundaries in the investigated steel resulted in a shear stress of approximately 1--1.5 GPa (shear modulus of austenite being 65 GPa \cite{Allain2004a}), significantly larger than the shear stress needed to retard twinning in $\alpha$-uranium. This large shear stress may help to explain why deformation twins in TWIP steels remain very thin. 

However, it should be noted that the twins imaged in the current sample were only 4 nm thick, whereas the \enquote{saturation} twin thickness in TWIP steels is approximately 30 nm \cite{DeCooman2018}, suggesting that these twins will continue to thicken with increasing strain. Based on the conclusions by Grilli \textit{et al.} \cite{Grilli2020} that twin growth is stopped when a critical density of residual dislocations on the twin boundary is reached, it is speculated that by the time the twin has grown to approximately 30 nm, the strain field along the twin boundary would have become more uniform, \textit{i.e.} no more hot spots. The high density of dislocations along the twin boundary would then be able to completely stop the motion of Shockley partial dislocations along the twin boundary and the twin would then reach its maximum thickness.


The existence of hot spots may also help to explain the variability of twin thickness. De Cooman \textit{et al.} \cite{DeCooman2018} showed that while twins are commonly considered to be plate-like in morphology, TEM observations have shown twin thicknesses to be highly variable and twin boundaries to be highly faulted \cite{Dao2006}. The formation of faulted twin boundaries appear to be caused by the disorderly overlapping of Shockley partial dislocations \cite{DeCooman2018}. Therefore, it is speculated that while the hot spots may not have been sufficient to completely stop twin thickening at early strains, they will certainly disrupt the propagation of Shockley partials along the twin boundary leading to twin growth with irregular thicknesses. 










In summary, the 4D-STEM technique was used to experimentally observe the strain fields in and around several deformation twins in a TWIP steel with sub-nm resolution. The results show several strain hot spots of up to 12\% along the twin boundaries which very likely correspond to sessile Frank dislocations along the twin boundaries according to the \enquote{pole $+$ deviation} mechanism proposed by Idrissi and Schryvers \cite{Idrissi2012}. The observed average strain of 6\% corresponds to a large shear stress of 1--1.5 GPa along the twin boundaries compared to other materials and is likely to be responsible for early retardation of twin growth and thickness saturation. This mechanism for twin thickness saturation helps to keep deformation twins in TWIP steels extremely fine, resulting in increased grain fragmentation with applied strain which is central to the dynamic Hall-Petch effect and underpins the impressive tensile properties of TWIP steels.

\vspace{\baselineskip}
\noindent
\textbf{Acknowledgements}
\vspace{\baselineskip}

TPM and DD would like to acknowledge support from the Rolls-Royce plc - EPSRC Strategic Partnership in Structural Metallic Systems for Gas Turbines (EP/M005607/1), and the Centre for Doctoral Training in Advanced Characterisation of Materials (EP/L015277/1) at Imperial College London. TWJK is grateful for a studentship from A*STAR. AA acknowledges EPSRC grant IAA EP/R511547/1. We thank Diamond Light source for access and support in use of the ePSIC instrument E02 and proposal no. EM18770 that contributed to the results presented here. BHS and CO acknowledge funding from the Toyota Research Institute, and from by the Office of Science, Office of Basic Energy Sciences, of the U.S. Department of Energy under contract no. DE-AC02-05CH11231.  We are grateful to TB Britton for insight into electron diffraction and micromechanics and B Poole for guidance with basis rotation. We also thank VA Vorontsov and AJ Knowles for helpful conversations when planning the project. 

\vspace{\baselineskip}
\noindent
\textbf{References}
\vspace{\baselineskip}

\bibliographystyle{thesis_thomas5}
\bibliography{Library}

\end{document}